\documentclass{vgtc}

\graphicspath{{figures/}{pictures/}{images/}{./}} % where to search for the images

\usepackage{times}                     % we use Times as the main font
\usepackage{booktabs}         
\usepackage{mathptmx}                  % use matching math font

\usepackage{glossaries}
\usepackage[inline]{enumitem}
\usepackage{tabularx} 
\usepackage{multirow}
\usepackage{array}

\newcolumntype{Y}{>{\centering\arraybackslash}X}
\usepackage{array}
\usepackage[nolist]{acronym}
\newcommand{\pval}[3]{%
  \ensuremath{#2\;\vcenter{\hbox{\scalebox{0.65}{$\begin{array}{@{}c@{}} #1 \\[-0.3ex] #3 \end{array}$}}}}%
}

\vgtcinsertpkg

\title{Hybrid Foveated Path Tracing with Peripheral\\Gaussians for Immersive Anatomy}

\author{Constantin Kleinbeck \thanks{\,Corresponding author, e-mail: constantin.kleinbeck@tum.de. \\Scheduled for publication in the Proceedings of IEEE VR 2026.}{\,\,$^{,}$}\textsuperscript{1, 2, 3, 4}\\ %
\and Luisa Theelke\textsuperscript{1, 2, 3, 4}\\ %
\and Hannah Schieber\textsuperscript{1, 2, 3, 4}\\ %
\and Ulrich Eck\textsuperscript{1}\\ %
\and Rüdiger von Eisenhart-Rothe\textsuperscript{1, 2, 4}\\ %
\and Daniel Roth\textsuperscript{1, 2, 3, 4}}
\affiliation{%
\scriptsize \textsuperscript{1} Technical University of Munich (TUM), Germany\\
\scriptsize \textsuperscript{2} TUM University Hospital, Orthopedics and Sports Orthopedics\\
\scriptsize \textsuperscript{3} Human-Centered Computing and Extended Reality Lab (HEX)\\ 
\scriptsize \textsuperscript{4} Munich Institute of Robotics and Machine Intelligence (MIRMI)}

\teaser{
  \centering
  \includegraphics[width=\linewidth]{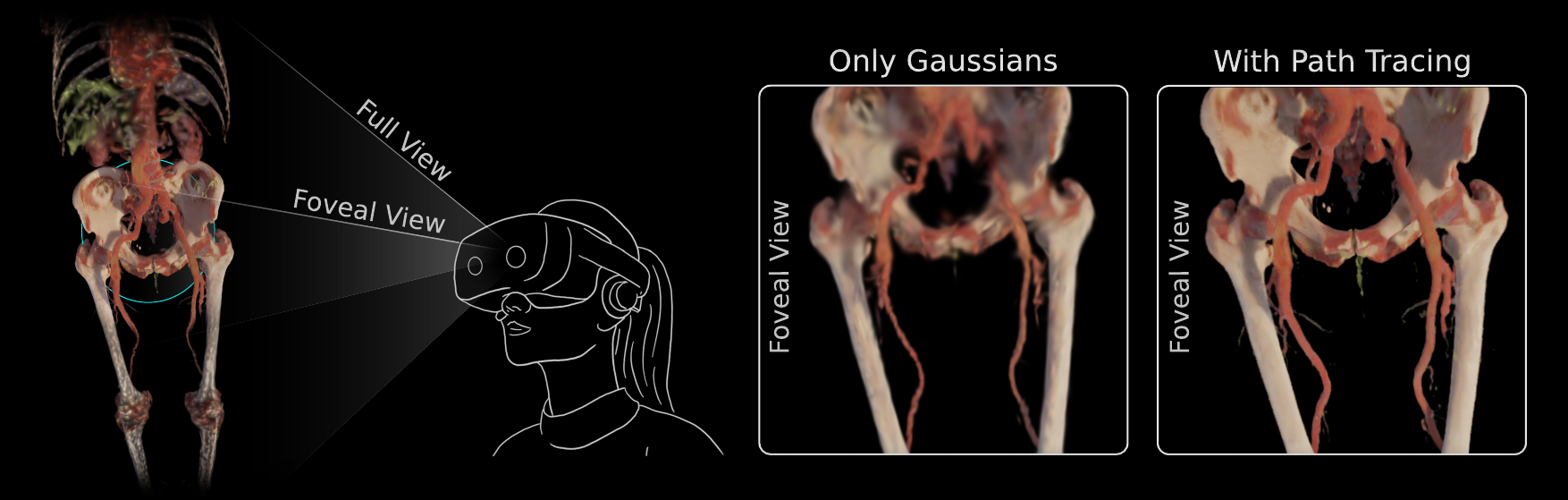}
  \caption{Our approach combines path tracing for the foveated region with a rapidly generated Gaussian Splatting model for the periphery. This combination utilizes the strengths of both approaches and enables anatomy rendering at a higher quality than just path tracing while overcoming the limited interactivity of only using Gaussians. The left side shows stylized Gaussians with overlay path tracing, the right close-up views of the foveal region, without and with combined path tracing.}
  \label{fig:teaser}
}

\abstract{
Volumetric medical imaging offers great potential for understanding complex pathologies. Yet, traditional 2D slices provide little support for interpreting spatial relationships, forcing users to mentally reconstruct anatomy into three dimensions. Direct volumetric path tracing and VR rendering can improve perception but are computationally expensive, while precomputed representations, like Gaussian Splatting, require planning ahead. Both approaches limit interactive use.

We propose a hybrid rendering approach for high-quality, interactive, and immersive anatomical visualization. Our method combines streamed foveated path tracing with a lightweight Gaussian Splatting approximation of the periphery. The peripheral model generation is optimized with volume data and continuously refined using foveal renderings, enabling interactive updates. Depth-guided reprojection further improves robustness to latency and allows users to balance fidelity with refresh rate.
We compare our method against direct path tracing and Gaussian Splatting. 
Our results highlight how their combination can preserve strengths in visual quality while re-generating the peripheral model in under a second, eliminating extensive preprocessing and approximations. This opens new options for interactive medical visualization.
} % end of abstract


\nocopyrightspace

\begin{document}

%%% Acro

\begin{acronym}[Bspwwww.]
%Az
\acro{ar}[AR]{augmented reality}
\acro{ap}[AP]{average precision}
\acro{api}[API]{application programming interface}
\acroplural{ann}[ANN]{artifical neural networks}
%B
\acro{bev}[BEV]{bird eye view}
\acro{rbob}[BRB]{Bottleneck residual block}
\acroplural{rbob}[BRBs]{Bottleneck residual blocks}
\acro{mbiou}[mBIoU]{mean Boundary Intersection over Union}
%C
% \textit{Cross Entropy Loss}
\acro{cai}[CAI]{computer-assisted intervention}
\acro{ce}[CE]{cross entropy}
\acro{cad}[CAD]{computer-aided design}
\acro{cnn}[CNN]{convolutional neural network}
\acro{ct}[CT]{computed tomography}
\acro{crf}[CRF]{conditional random fields}
%D dla
\acro{dpc}[DPC]{dense prediction cells}
\acro{dla}[DLA]{deep layer aggregation}
\acro{dnn}[DNN]{deep neural network}
\acroplural{dnn}[DNNs]{deep neural networks}
\acro{dvr}[DVR]{direct volume rendering}
\acro{da}[DA]{domain adaption}
\acro{dr}[DR]{domain randomization}
%E
%F
\acro{fat}[FAT]{falling things}
\acro{fcn}[FCN]{fully convolutional network}
\acroplural{fcn}[FCNs]{fully convolutional networks}
\acro{fov}[FoV]{field of view}
\acro{fv}[FV]{front view}
\acro{fp}[FP]{False Positive}
\acro{fpn}[FPN]{feature Pyramid network}
\acro{fn}[FN]{False Negative}
\acro{fmss}[FMSS]{fast motion sickness scale}
%G
\acro{gan}[GAN]{generative adversarial network}
\acroplural{gan}[GANs]{generative adversarial networks}
\acro{gcn}[GCN]{graph convolutional network}
\acroplural{gcn}[GCNs]{graph convolutional networks}
\acro{gs}[3DGS]{3D Gaussian Splatting}
%H
\acro{hmi}[HMI]{Human-Machine-Interaction}
\acro{hmd}[HMD]{Head Mounted Display}
\acroplural{hmd}[HMDs]{head mounted displays}
%I
\acro{iou}[IoU]{intersection over union}
\acro{irb}[IRB]{inverted residual bock}
\acroplural{irb}[IRBs]{inverted residual blocks}
\acro{ipq}[IPQ]{igroup presence questionnaire}
%J
%K
\acro{knn}[KNN]{k-nearest-neighbor}
%L
\acro{lidar}[LiDAR]{light detection and ranging}
\acro{lsfe}[LSFE]{large scale feature extractor}
\acro{llm}[LLM]{large language model}
%M
\acro{map}[mAP]{mean average precision}
\acro{mc}[MC]{mismatch correction module}
\acro{miou}[mIoU]{mean intersection over union}
\acro{mis}[MIS]{Minimally Invasive Surgery}
\acro{msdl}[MSDL]{Multi-Scale Dice Loss}
\acro{ml}[ML]{Machine Learning}
\acro{mlp}[MLP]{multilayer perception}
\acro{miou}[mIoU]{mean Intersection over Union}
\acro{mri}[MRI]{magnetic resonance imaging}
%N neural networks
\acro{nn}[NN]{neural network}
\acroplural{nn}[NNs]{neural networks}
\acro{ndd}[NDDS]{NVIDIA Deep Learning Data Synthesizer}
\acro{nocs}[NOCS]{Normalized Object Coordiante Space}
\acro{nerf}[NeRF]{Neural Radiance Fields}
\acro{NVISII}[NVISII]{NVIDIA Scene Imaging Interface}
\acro{ngp}[NGP]{neural graphics primitives}
%O
\acro{or}[OR]{Operating Room}
%P
\acro{pbr}[PBR]{physically based rendering}
\acro{psnr}[PSNR]{peak signal-to-noise ratio}
\acro{pnp}[PnP]{Perspective-n-Point}
%Q
%R
\acro{rv}[RV]{range view}
\acro{roi}[ROI]{region of interest}
\acroplural{roi}[ROIs]{region of interests}
\acro{rbab}[BB]{residual basic block}
\acro{ras}[RAS]{robot-assisted surgery}
\acroplural{rbab}[BBs]{residual basic blocks}
%S
\acro{spp}[SPP]{samples per pixel}
\acro{sh}[SH]{spherical harmonics}
\acro{sgd}[SGD]{stochastic gradient descent}
\acro{sdf}[SDF]{signed distance function}
\acro{sfm}[SfM]{structure-from-motion}
\acro{sam}[SAM]{Segment-Anything}
\acro{sus}[SUS]{system usability scale}
\acro{ssim}[SSIM]{structural similarity index measure}
\acro{sfm}[SfM]{structure from motion}
\acro{slam}[SLAM]{simultaneous localization and mapping}
%T
\acro{tp}[TP]{True Positive}
\acro{tn}[TN]{True Negative}
\acro{thor}[thor]{The House Of inteRactions}
\acro{tsdf}[TSDF]{signed distance function}
%U
%V
\acro{vr}[VR]{Virtual Reality}
%W
\acro{ycb}[YCB]{Yale-CMU-Berkeley}

\acro{ar}[AR]{augmented reality}
\acro{ate}[ATE]{absolute trajectory error}
\acro{bvip}[BVIP]{blind or visually impaired people}
% C
\acro{cnn}[CNN]{convolutional neural network}
\acro{c2f}[c2f]{coarse-to-fine}
%F
\acro{fov}[FoV]{field of view}
%G
\acro{gan}[GAN]{generative adversarial network}
\acro{gcn}[GCN]{graph convolutional Network}
\acro{gnn}[GNN]{Graph Neural Network}
%H
\acro{hmi}[HMI]{Human-Machine-Interaction}
\acro{hmd}[HMD]{head-mounted display}
\acro{mr}[MR]{mixed reality}
% I
\acro{iot}[IoT]{internet of things}
% L
\acro{llff}[LLFF]{Local Light Field Fusion}
\acro{bleff}[BLEFF]{Blender Forward Facing}

\acro{lpips}[LPIPS]{learned perceptual image patch similarity}
%N
\acro{nerf}[NeRF]{neural radiance fields}
\acro{nvs}[NVS]{novel view synthesis}
% M
\acro{mlp}[MLP]{multilayer perceptron}
\acro{mrs}[MRS]{Mixed Region Sampling}

%O
\acro{or}[OR]{Operating Room}
%P
\acro{pbr}[PBR]{physically based rendering}
\acro{psnr}[PSNR]{peak signal-to-noise ratio}
\acro{pnp}[PnP]{Perspective-n-Point}
%Q
%R
%
\acro{sus}[SUS]{system usability scale}
\acro{ssim}[SSIM]{similarity index measure}
\acro{sfm}[SfM]{structure from motion}
\acro{slam}[SLAM]{simultaneous localization and mapping}

%T
\acro{tp}[TP]{True Positive}
\acro{tn}[TN]{True Negative}
\acro{thor}[thor]{The House Of inteRactions}
%U
\acro{ueq}[UEQ]{User Experience Questionnaire}
%V
\acro{vr}[VR]{virtual reality}
%W
\acro{who}[WHO]{World Health Organization}
%X
\acro{xr}[XR]{extended reality}
%Y
\acro{ycb}[YCB]{Yale-CMU-Berkeley}
\acro{yolo}[YOLO]{you only look once}
\end{acronym}

%%% Paper

\firstsection{Introduction}

\maketitle

Visualizing medical anatomy holds significant potential for education, diagnosis, treatment planning, and intervention \cite{liang_rewview_2022}. Although clinical imaging modalities such as \ac{ct} or \ac{mri} inherently capture the three-dimensional structure of the body, most visualization workflows for diagnosis and planning still present these data as 2D sequential cross-sectional slices. Consequently, users are required to mentally reconstruct these 2D images into a 3D understanding of human anatomy. This cognitive transformation is time-consuming and challenging, especially for students, patients, and other non-experts \cite{battulga_effectiveness_2012}, and could be significantly alleviated by techniques that allow the 3D anatomy to be directly visualized and explored.

Advances in rendering have pushed the boundaries of medical visualizations for many years, bringing more advanced ways to present anatomical data~\cite{calhoun_three-dimensional_1999}. In particular, 3D visualization techniques have been adapted for anatomical data, aiming to present complex structures in a more intuitive and accessible way \cite{liang_rewview_2022}. Approaches range from direct volume rendering \cite{engel_real-time_2004} to volumetric path tracing \cite{caton_jr_three-dimensional_2020}, improving visual quality, especially in rendering of lighting and material interactions. These, in turn, aid in understanding complex spatial relationships and have shown success in medical education~\cite{binder_cinematic_2021, cardobi_path_2023}, as well as clinical analysis, and planning~\cite{rowe_application_2019, elshafei_comparison_2019}. Recently, intermediate point-based representations like \ac{gs} \cite{kerbl_3d_2023} have been used to improve rendering speed while sacrificing interactivity, enabling viewers to crop or adjust visibilities to focus on certain areas~\cite{niedermayr_application_2024, kleinbeck_multi-layer_2025}. Each approach brings with it distinct trade-offs between realism, interactivity, and computational costs. 

Although 3D visualization techniques make anatomical structures perceptible, most medical applications still rely on 2D visualization, limiting depth perception and interactivity. Immersive displays such as \ac{vr} headsets overcome these limitations, providing stereoscopic cues and head-tracked exploration that improve comprehension of complex anatomical relationships~\cite{luijten_screen_2025}. High-quality volumetric path tracing in \ac{vr}, however, remains prohibitively expensive at the refresh rates and resolutions required for comfortable, low-latency immersion. Intermediate representations, such as polygon meshes or point clouds, can accelerate rendering, but they sacrifice fine anatomical detail and limit flexibility in modifying appearance interactively. Perceptual optimization strategies, particularly foveated rendering, exploit the eye’s nonuniform acuity to allocate rendering resources where they matter most \cite{weier_perception-driven_2017}. Recent works further indicate that the improved depth understanding provided by \ac{vr} is not affected by foveated rendering~\cite{kergasner_towards_2025}. Yet, scaling these techniques to the real-time performance targets of modern, high-resolution \ac{vr} displays remains a challenge~\cite{mohanto_integrative_2022}.

To address this problem, we propose a hybrid rendering system for immersive anatomical visualization that balances quality, speed, and interactivity. Our approach combines streamed high-quality, volumetric path tracing for the foveal region with a rapidly generated \ac{gs} for peripheral representation, rendered at native headset refresh rates. By restricting computationally intensive path tracing for the user’s foveal region, we concentrate rendering resources where perceptual sensitivity is highest. Representing peripheral regions with a rapidly generated approximation provides a visually plausible anatomy at a fraction of the computational cost. This peripheral model can be created in seconds and continuously refined using streamed foveal renderings, improving in quality as the user explores the volume. In addition, by guiding reprojection with volume depth, we decouple render speed from display refresh rate, providing a controllable latency–quality trade-off. 

This hybrid strategy combines the strengths of both foveated, high-fidelity path tracing and rapid Gaussian-based peripheral rendering. Compared to pure path tracing, it substantially improves real-time performance and foveal rendering quality. Compared to fully precomputed intermediate representations, it reduces initial setup time from minutes to seconds. It further enables interactive operations such as transfer function changes or clipping planes by regenerating the model within seconds, and maintains the ability to display fine anatomical details beyond those present at the time of model creation. Notably, this approach integrates both path tracing rendering and \ac{nvs}, leveraging advancements in either domain while keeping its own strengths.

\textbf{Contribution:}
This work contributes the following:
\setlist[itemize]{itemsep=0.5em, topsep=0.5em, parsep=0pt}
\begin{itemize}
    \item A novel hybrid rendering pipeline for anatomy integrating streamed foveated path tracing and a regenerating \ac{gs} peripheral cloud model
    \item A depth-guided reprojection method for the path-traced anatomy to hide latency and decouple render speed from display refresh
    \item An evaluation exploring interactive peripheral model generation, system performance, and perceptual fidelity compared to standalone path tracing and \ac{gs}
\end{itemize}

\section{Related Work}
\label{sec:rel_work}

\begin{figure*}[t!]
 \centering %
 \includegraphics[width=\linewidth]{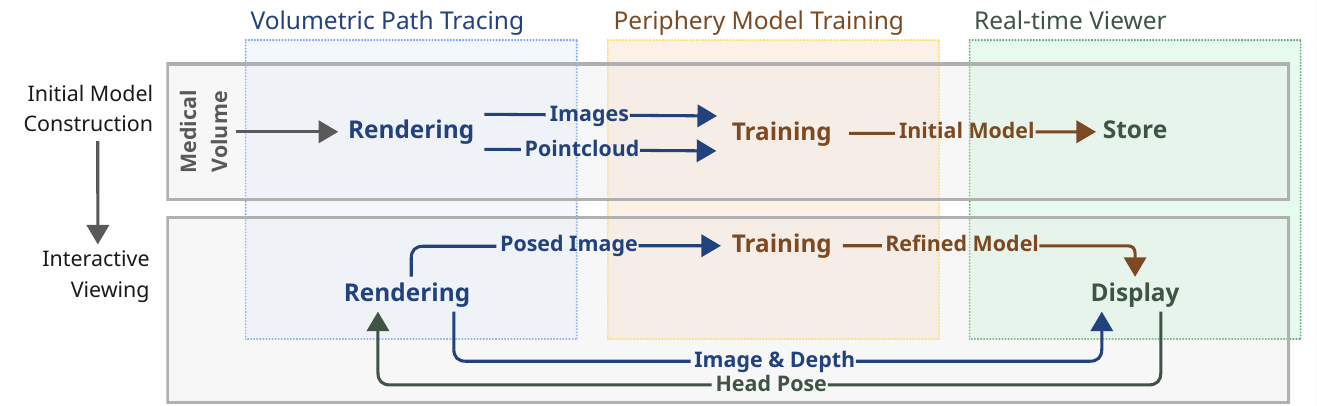}
 \caption{System design showing the three main components and the two application flows between them. The initial model construction is always executed before interactive viewing, which uses and improves the initial model with the same images originally rendered for the real-time viewer.}
 \label{fig:overview}
\end{figure*}

This work builds on three main research directions: volumetric path tracing for anatomical volumes, novel view synthesis for the peripheral model, and immersive rendering that brings these components to \ac{vr}. These three components are the major modules of our system, as illustrated in \autoref{fig:overview}, and structure the subsections below.

\subsection{Volumetric Path Tracing}

Rendering of volumetric data, usually called participating media, is a long-standing problem in graphics but remains an active area of research~\cite{novak_monte_2018}. Developed for general rendering, these methods were adapted to medical volumes from \ac{ct}, \ac{mri} scans as \ac{dvr}~\cite{zhang_volume_2011}.
Monte Carlo volumetric path tracing yields higher visual fidelity than previous marching approaches~\cite{kroes_exposure_2012, dappa_cinematic_2016}, at much higher computational cost and noise. Colorization of anatomy is commonly achieved through a user-adjustable transfer function, mapping volume density to colors and other attributes~\cite{ljung_state_2016}. For interactive and immersive applications, however, longer render times are not feasible, making denoising a crucial component. Consequently, substantial work targets denoising for volumetric data, including general~\cite{hofmann_joint_2023} and medical volumes~\cite{hofmann_neural_2020, iglesias-guitian_real-time_2022}, noting online medical volumetric denoising as an ongoing challenge due to noisy buffers. Multiple works investigate neural architectures~\cite{chaitanya_interactive_2017} as well as statistical and spatio-temporal approaches~\cite{schied_spatiotemporal_2017}, with a focus on speed. Some approaches are tuned specifically for \ac{vr} viewing, exploiting shared information between the eyes~\cite{taibo_immersive_2024}. Our pipeline must denoise both disjoint initial renders not targeting \ac{vr}, as well as temporally accumulated frames. Thus, we rely on a fast deep-learning denoiser.

\subsection{Novel View Synthesis}

\Acl{nvs} (\acs{nvs}) refers to techniques that generate unseen viewpoints from a limited set of input images. In particular, \ac{nvs} based on \ac{gs}~\cite{kerbl_3d_2023} produces a high-quality set of Gaussian primitives that can be rendered efficiently with established graphics primitives. While training can start from noise, it is typically initialized using pre-existing point clouds, which are then densified. Several works propose deriving the initial point cloud directly from images to reduce training time and improve robustness with sparse views. Depth estimation~\cite{chen_mvsplat_2025}, 3D convolutions~\cite{liu_mvsgaussian_2024}, or pretrained transformers~\cite{fan_instantsplat_2025, jiang_anysplat_2025} have seen use. Although these strategies accelerate optimization, the overhead of estimating the initialization often makes them impractical for on-the-fly anatomical reconstructions, with pretrained models challenged by anatomical data.

Beyond initialization, recent research addresses runtime efficiency. 3DGS.zip provides an extensive overview~\cite{bagdasarian_3dgszip_2025}. While compression of trained splats is a central theme, other directions focus on reducing Gaussian creation or improving optimization efficiency itself. Notable contributions include hierarchical tree structures~\cite{ren_octree-gs_2025}, more targeted densification~\cite{mallick-taming_2024}, and simplification methods~\cite{fang_mini-splatting_2024}. Mini-Splatting2~\cite{fang_mini-splatting2_2024} improves optimization and integrates revised rasterization techniques~\cite{mallick-taming_2024}. It bases its simplification on accurate Gaussian positions and targets a low number of final Gaussians. As we can generate precisely placed initial points from volumes and require a lightweight peripheral model, we build our approach upon this work. Originally developed for static scenes, subsequent research has extended \ac{gs} to dynamic or incremental settings, enabling continual integration of new views with localized optimization~\cite{ackermann_cl-splats_2025}, real-time on-the-fly scene optimization~\cite{meuleman_onthefly_2025}, streaming applications~\cite{yan_instant_2025}, and even SLAM~\cite{matsuki_gaussian_2024}. These approaches further inform continued optimization use cases.

\ac{gs} has also been adopted for medical applications. Multiple works investigate \ac{gs} from endoscopic video~\cite{li_endosparse_2024}, possibly in real-time. In anatomical visualization, \ac{gs} offers faster rendering compared to path tracing by shifting computation into a preprocessing step~\cite{niedermayr_application_2024}. Pretrained encoder–decoder networks can enable rapid initialization~\cite{gao_render-fm_2025} given volume and color data, similar to view-based predictions. Further extensions support interactive visualization through clipping-aware techniques~\cite{li_clipgs_2025} and multi-layer rendering~\cite{kleinbeck_multi-layer_2025}. While being able to display high-quality renderings in \ac{vr} at high frame rates with some interactivity, the methods lack the complete control that direct path tracing provides.

\subsection{Immersive Rendering}

A central challenge in immersive visualization is rendering high-resolution stereo views at high frame rates. An approach to render efficiently is foveated rendering~\cite{patney_towards_2016}. It operates by concentrating computational resources and thus visual quality in the foveal region, where visual acuity is highest: roughly within the central 5° of vision, tapering off to around 17°~\cite{mohanto_integrative_2022}. Beyond focusing on rendering, works have also explored simplified peripheral scene regions to reduce computational demands. Examples include the use of proxy geometries~\cite{waldow_investigating_2024}, typically pre-generated and not refined over time. To meet strict frame-rate requirements, reprojection techniques reuse previously rendered information~\cite{van_waveren_asynchronous_2016}, and can be adapted to specific applications like in foveated ray tracing pipelines~\cite{weier_foveated_2016}. However, issues like disocclusions caused by large viewpoint changes or overlapping content remain problematic~\cite{mark_post-rendering_1997}. High-speed eye-tracking is often used as a basis for foveated rendering, with studies showing that foveated rendering can tolerate delays of up to 50\,ms, much higher than the sub-10\,ms required for full-frame real-time rendering~\cite{albert_latency_2017}. This tolerance motivates our exploration of slower update rates for foveated rendering, trading refresh speed for improved visual quality. Another way to support resource-constrained devices are split-rendering architectures, distributing workloads between a server and client. These systems may stream shading data~\cite{mueller_shading_2018}, scene representations~\cite{hladky_quadstream_2022}, high-quality lighting~\cite{stengel_distributed_2021}, or partial views for local composition~\cite{lim_adaptive_2024}. Our system aligns with this research, using a split-rendering approach to maximize the quality of streamed anatomical views.

Several works combine these strategies with \ac{gs}. In \ac{vr}, \ac{gs} has been applied to various visualization tasks~\cite{li_radiance_2025}, sometimes enhanced with additional data such as semantic labels~\cite{schieber_semantics-controlled_2025}. Especially relevant are studies integrating foveated rendering with \ac{gs} for efficiency~\cite{lin_metasapiens_2025}. VRsplat~\cite{tu_vrsplat_2025} combines foveated Gaussian rendering with improved projection techniques~\cite{huang_error_2025} and artifact prevention~\cite{radl_stopthepop_2024}. VR-Splatting~\cite{franke_vr-splatting_2024} extends this idea by combining Gaussians for peripheral vision with higher-quality point-based rendering in the fovea, closely resembling our strategy. Closest to our work, Henriques et al.~\cite{henriques_foveated_2024} combine foveated path tracing with a peripheral \ac{gs} model for mesh-based scenes. However, this work targets general scenes and does not address the on-the-fly regeneration of the peripheral model or use guided reprojection to mitigate latency, which are key contributions of our approach.

\subsection{Research Gap}

There is a clear need for high-quality anatomical rendering in immersive environments. Path-traced methods now provide reasonably good results in \ac{vr}, but maintaining consistent performance, particularly on standalone mobile devices, remains challenging. Streaming path tracing at high and steady frame rates is difficult, with limited options to handle latency. \ac{gs} has been proposed as an intermediate representation, offering speed advantages, but existing approaches fall short of providing full interactivity. Our work addresses this gap by enabling high-quality, fully interactive anatomical rendering on mobile devices, while incorporating latency masking techniques to improve user experience.

\begin{figure*}[t!]
 \centering %
 \includegraphics[trim={2.6cm 0cm 4.2cm 0cm},clip,width=\linewidth]{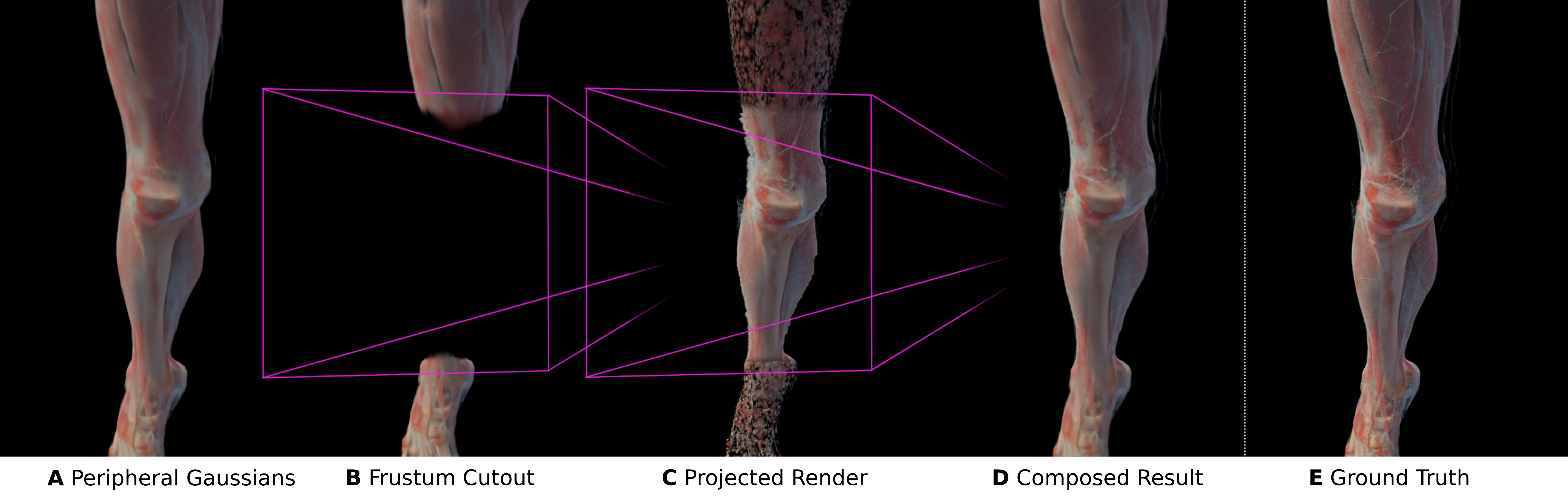}
 \caption{Composition process of the foveated and peripheral region, as performed in the real-time viewer. \textbf{A} shows peripheral Gaussians, \textbf{B} how Gaussians are discarded in a downscaled frustum shape to enable correct transparency in the rendered views. \textbf{C} shows how the foveal view is composited at depth while being blended at the edges of the foveal view. Steps \textbf{B} and \textbf{C} show slightly angled views, with the drawn frustum highlighting the original camera view. In \textbf{C}, gaussians are downscaled for visual clarity in this image.}
 \label{fig:visual_components}
\end{figure*}

\section{System Design}
\label{sec:sys_design}

Our system interactively renders volumetric medical scans by combining the strengths of foveated path tracing with a peripheral model based on \ac{gs}. This hybrid approach focuses computational resources on path tracing the user's foveal region at high quality, while the faster \ac{gs} model provides the peripheral view. To maintain interactivity when users modify rendering parameters, such as colorization or cutting planes, the peripheral \ac{gs} model is designed to be regenerated within seconds. An objective is to work with consumer-level hardware and mobile \ac{vr} devices.

The system's components and workflow are illustrated in \autoref{fig:overview}. Architecturally, it consists of three independent components:
(i) a lightweight \ac{vr} real-time viewer, (ii) a GPU-intensive volumetric path tracer, and (iii) a \ac{gs} peripheral model training module with an irregular, GPU-intensive workload.
This modularity allows the subsystems to be independent of physical machines. With drawbacks in render and peripheral model generation speed, they could share computing resources. It operates in two modes: (i) For initial model construction, the path tracer generates posed images based on the viewer's render settings, from which the \ac{gs} component creates an initial peripheral model. (ii) During interactive use, the real-time viewer sends the current head pose to the path tracer, which returns a high-quality foveated image. In the viewer, this image is depth-projected and composited with the peripheral \ac{gs} model, as shown in \autoref{fig:visual_components}. Crucially, these same foveated images are also fed to the \ac{gs} optimization subsystem, allowing for continuous refinement of the peripheral model with no additional image generation overhead. The improved \ac{gs} model is then periodically sent back to the viewer, replacing the previous one. This cycle repeats until a change in render settings triggers a full regeneration of the model. Conceptually, the path tracer and training module run on a high-power remote machine, while the lightweight viewer can run on any modern desktop machine or standalone \ac{vr} headset with DirectX 12 or Vulkan support.

\section{Method}

As outlined in \autoref{sec:sys_design}, our system is comprised of three main components. The following sections describe each component and its interactions in detail.

\subsection{Volumetric Path Tracing}

Our volumetric path tracer generates both the initial image set for training the peripheral model and the continuous foveated views for the user. It processes DICOM data using a supplied transfer function and an environment map for lighting. We opted to use a limited and focused path tracer for ease of adaptation to our needs, however, the approach works with any path tracer providing posed images in response to queries. The primary trade-off between quality and performance is managed by the sample count per pixel (\acs{spp}), which we evaluate from 8 to 32. We limit the number of path bounces to a maximum of 4, as further bounces offered negligible quality improvements at a significant computational cost.

We integrate the NVIDIA OptiX denoiser~\cite{chaitanya_interactive_2017}, as it works well on standalone images as well as temporally correlated series. Initial images used for model generation, where viewpoints are diverse, are denoised individually. For interactive viewing, we apply temporal denoising using motion vectors and albedo buffers. Motion vectors are derived from the first significant hit point within the volume, a heuristic that has been shown to perform well for semi-transparent surfaces~\cite{iglesias-guitian_real-time_2022}. To generate a stable albedo, we perform a limited raymarching step from the hit point without environment lighting. This better approximates the volume's surface color than directly sampling the transfer function at a minor performance cost.

To create the initial point cloud, we cast rays towards the volume from various viewpoints around it. We then store the first significant hit point for each successful intersection with a volume surface. These points are subsequently colored using the same path tracing logic as our final images. This method ensures that the point cloud is well-aligned with the visible surface. The entire process is implemented in a compute shader, generating and coloring 20,000 points with 64 samples each in approximately 50\,ms. Fewer samples result in noisy colors, with the number of points empirically chosen as a reasonable starting point for optimization. We do not use any specular terms or other information besides RGBA color in our transfer functions.

\begin{figure}[t!]
 \centering %
 \includegraphics[trim={1.2cm 0cm 3.7cm 0cm},clip,width=\linewidth]{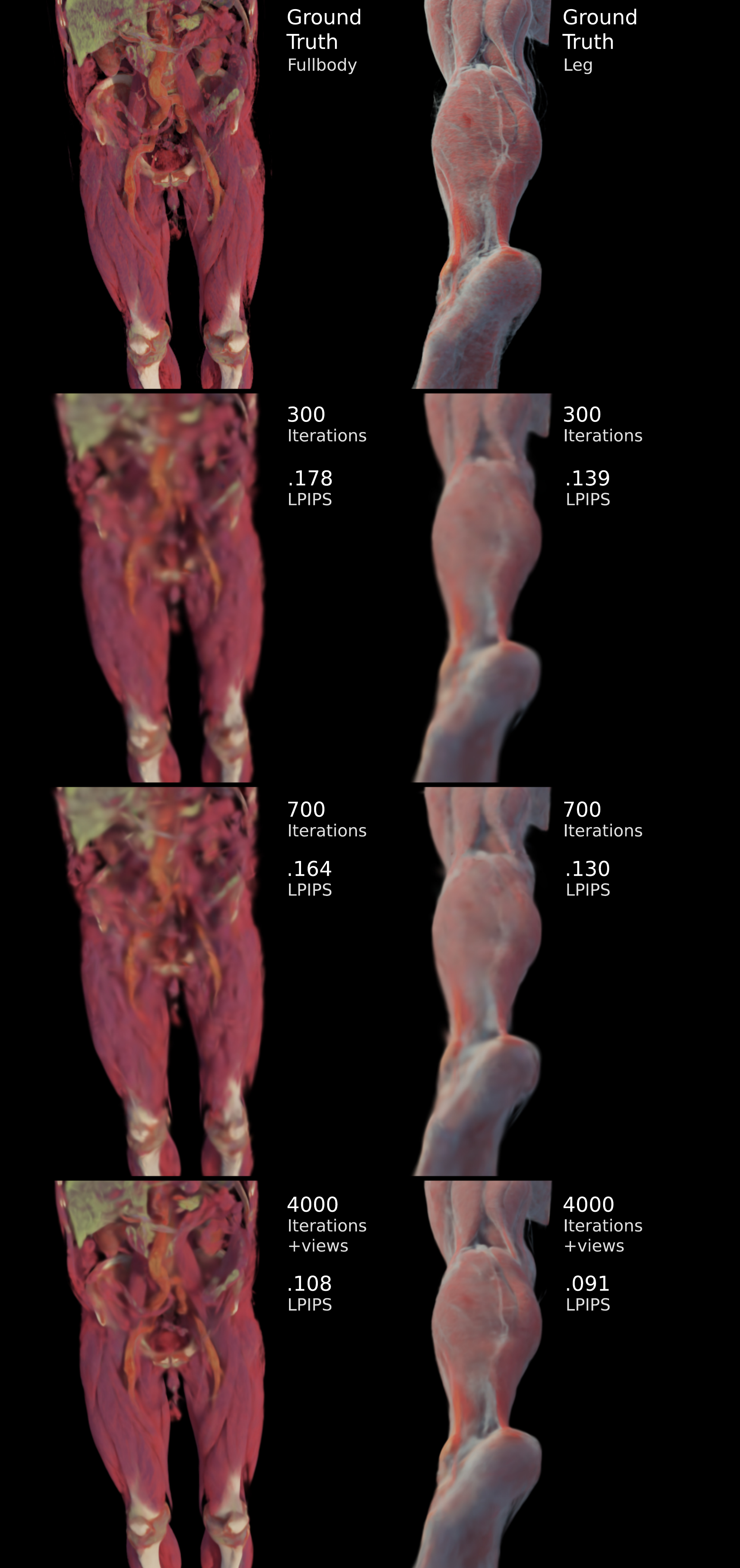}
 \caption{Visual overview of \emph{peripheral model initialization progress}. Shown are two scenes, initialized from 12 images with 8 spp. We further include the quality after continual training with 512 additional views with  4 spp each.}
 \label{fig:init_compare}
\end{figure}

\subsection{Periphery Model Training}
\label{ss:peripheralmodeltraining}

\begin{figure*}[t!]
 \centering %
 \includegraphics[width=\linewidth]{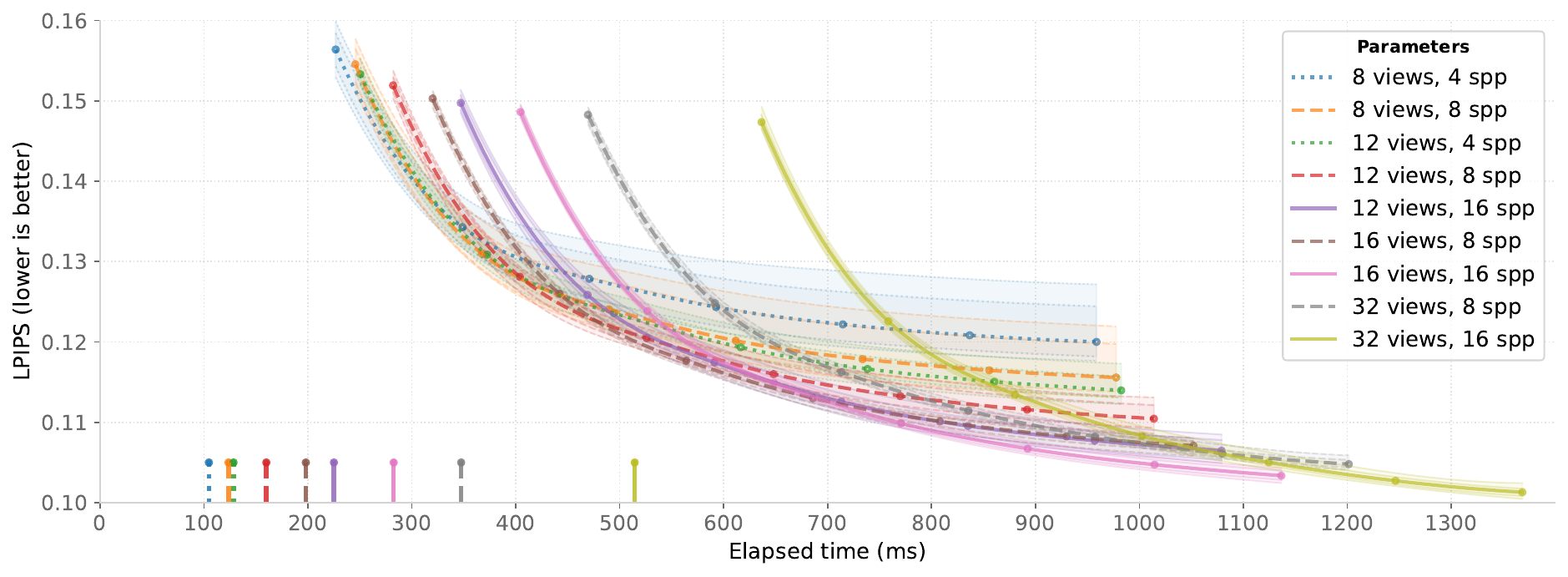}
 \caption{Comparison of multiple initialization numbers of views and samples per pixel. The time on the x-axis includes the time to render the views and generate the initial point cloud. The render end time is indicated by the vertical lines on the x-axis. This means that every plotted point represents the visual quality that users would see at that point in time in the viewer, without transmission and loading overhead. Each line contains average and maximum p10/p90 values from different datasets and random views, representing typical and best/worst quality bands.}
 \label{fig:init_params_comparison_graph}
\end{figure*}

The peripheral model is created in two phases: initial model construction and continuous refinement during viewing. We adapt Mini-Splatting2~\cite{fang_mini-splatting2_2024} as our basis for model optimization, augmented with techniques from Taming 3DGS~\cite{mallick-taming_2024}, for its rapid reconstruction capabilities. To better handle semi-transparent regions common in medical data, we incorporate alpha training and randomized backgrounds~\cite{niedermayr_application_2024, kleinbeck_multi-layer_2025}. Our surface-aligned initial point cloud is particularly well-suited to Mini-Splatting2's position-based simplification training strategy.

For initial peripheral model creation, we perform an abbreviated training of roughly one second on a small image set. We bypass the densification step and apply simplification after only 95 optimization steps. To accelerate convergence, we increase the learning rates for scale, opacity, and color by about 50\%, a value we determined empirically. This process rapidly produces a model of approximately 10,000 Gaussians. It also effectively prevents the formation of floater artifacts even with very sparse image sets, as no new Gaussians are created.

After initialization, the model is continuously refined using the foveated images generated during user interaction. These views often contain high-frequency details from close-up perspectives that were absent in the initial, wider-angle views. However, during prolonged use, many training images with redundant information are accumulated. To manage resources, we propose a simple heuristic whereby a new foveated view is added to the training set only if it provides sufficiently novel information:
\begin{align}
\forall C_i \in \text{Set}_{views}: \quad 
& (||pos_{new} - pos_i|| > \delta_{pos}) \nonumber \\
\lor\quad 
& (\vec{view}_{new} \cdot \vec{view}_i < \cos(\theta_{view}))
\end{align}
Here, the distance threshold $\delta_{pos}$ and the minimum angle threshold $\theta_{view}$ were empirically set to \textit{0.05} and \textit{5$^{\circ}$}, respectively, based on preliminary tests. For this work, we simplify the continual training to a periodic restart of the optimization process. Combining existing with new renders, we reload the existing Gaussians as initialization, but discarding the optimizer state.

\subsection{Real-time Viewer}

The viewer is built in Unity for broad compatibility with \ac{vr} devices and is compatible with tethered or standalone devices. It receives foveated images with corresponding depth data from the path tracer. Additionally, it receives initial and updated \ac{gs} models, however updating of the models is a manual process in this prototype. The \ac{gs} models are rendered using a publicly available plugin\footnote{\url{https://github.com/aras-p/UnityGaussianSplatting}}, with enhancements from recent work, such as view-center dependent sorting frequency~\cite{kleinbeck_multi-layer_2025}. We further adjust render settings to correct for inaccuracies during training. Increasing the splat size and opacity helps compensate for the systematic underestimation of the size of small structures like vessels in front of the background by the Gaussian optimization on low-resolution images. It also helps with a lack of visual brightness, likely also due to the low resolution and sample count in training images. Experimentation showed an increase of 10\% to be a good starting point, after which visual clarity and quality decrease.

To mask the latency of the path tracer, we project the received foveated image using the head position from when it was scheduled. This reprojection is guided by the depth data from the path tracer's first-hit points, which we compute anyway as a free byproduct of path tracing, but takes place entirely in a shader inside the real-time viewer. We found this to be more robust than estimating depth directly from the peripheral model. The original world-space position $p_{world}$ is reconstructed from the pixel's UV coordinates $uv \in [0,1]$ and its linear depth $d$ (world distance), where $P$ and $V$ are the projection and view matrices, and $\text{pos}_{old}$ is the camera position:
\begin{align}
p_{\text{far}} &= \frac{(P_{\text{old}}V_{\text{old}})^{-1} \,(2uv-1,\,1,\,1)^{\top}}{w},
\\
p_{\text{world}} &= \text{pos}_{\text{old}} + \frac{p_{\text{far}}-\text{pos}_{\text{old}}}{\|p_{\text{far}}-\text{pos}_{\text{old}}\|}\, d, \label{eq:reprojection}
\\
p'_{\text{clip}} &= P_{\text{new}}V_{\text{new}}\, p_{\text{world}}
\end{align}

The foveated image is projected onto a mesh. This mesh is dynamically deformed by the depth accompanying the foveated image and persists in the scene until the next foveated render arrives. This ensures stable, high-frame-rate rendering where the central view is updated asynchronously. We apply additional blending to smooth the transition between path traced image and \ac{gs} cloud. Minor disocclusions due to head movement are filled by the peripheral \ac{gs} model, providing coverage of newly visible areas~\cite{mark_post-rendering_1997}. A frustum-shaped cutout is applied to the \ac{gs} model where the foveated image is displayed, preventing incorrect blending with transparent regions in the streamed image.

\subsection{Interconnections}

A low-latency connection between the path tracer and the viewer is critical for human use, however, in this work we use a straightforward implementation favoring simplicity as proof of principle. We use a direct TCP connection and message passing between systems for communication, with MessagePack\footnote{\url{https://msgpack.org/index.html}} for serialization, selected for its performance and portability. Thus, system components can easily be spread out over multiple machines or mobile head-mounted devices, as long as a network connection is available. Furthermore, this setup lets us easily test components locally with connection via the loopback interface. Image textures are transmitted in their native format. This scheme has reasonable complexity and portability while providing a usable baseline experience. Further optimizing this network layer could be achieved by instead encoding and transmitting a low latency video stream, but is not a focus of this work. Real-time streaming and updating of improved peripheral models are not implemented in this prototype but were done manually for testing and evaluations.

\begin{table*}[t!]
\centering
\caption{Continual peripheral model refinement comparison. The Initial row shows data for the model before training with extension images.  The initial model quality has little influence on the final quality. Improvements after 2000 iterations are minor, more visible in perceptual metrics.}
\label{tab:ext-compare}
\begin{tabularx}{\linewidth}{@{}ll|YYYYYYll@{}}
\toprule
 &                                                     & \multicolumn{2}{c}{Initial} & \multicolumn{2}{c}{Early (2k)} & \multicolumn{2}{c}{Late (4k)} & \multicolumn{2}{c}{Gaussians} \\ 
\cmidrule(l){3-4} \cmidrule(l){5-6} \cmidrule(l){7-8} \cmidrule(l){9-10}
Init                                                                   & Extension        &  MPSNR                    & LPIPS                   & MPSNR                   & LPIPS                   & MPSNR                     & LPIPS                   & Init            & End          \\ \midrule
\multirow{2}{*}{\begin{tabular}[c]{@{}c@{}}$12 \times 8$ \end{tabular}}& $64 \times 4$    & \pval{18.17}{22.29}{25.59} & \pval{.073}{.110}{.163} & \pval{22.47}{25.82}{28.37} & \pval{.055}{.084}{.116} & \pval{22.15}{25.61}{28.21} & \pval{.054}{.081}{.111} & \pval{8k}{10k}{11k} &  \pval{17k}{20k}{23k} \\[2pt] % Add some slight separation between rows
    & $512 \times 16$  & \pval{18.17}{22.29}{25.59} & \pval{.073}{.110}{.163} & \pval{24.23}{26.96}{29.35} & \pval{.037}{.066}{.098} & \pval{24.50}{27.57}{30.00} & \pval{.034}{.062}{.091} & \pval{8k}{10k}{11k} &  \pval{28k}{35k}{46k} \\[3pt]
\multirow{2}{*}{\begin{tabular}[c]{@{}c@{}}$16\times16$ \end{tabular}} & $64 \times 4$    & \pval{18.45}{22.67}{26.24} & \pval{.067}{.103}{.155} & \pval{22.51}{25.84}{28.43} & \pval{.055}{.083}{.115} & \pval{22.08}{25.62}{28.31} & \pval{.053}{.081}{.111} & \pval{8k}{10k}{12k} & \pval{17k}{21k}{23k} \\[2pt]
    & $512 \times 16$  & \pval{18.45}{22.67}{26.24} & \pval{.067}{.103}{.155} & \pval{24.21}{27.01}{29.42} & \pval{.037}{.066}{.098} & \pval{24.59}{27.60}{30.05} & \pval{.034}{.061}{.091} & \pval{8k}{10k}{12k} & \pval{28k}{36k}{46k} \\ \bottomrule
\end{tabularx}
\begin{minipage}{\linewidth}
\vspace{2mm}
{\raggedright \scriptsize{\textit{Note:} MPSNR denotes masked PSNR values. Values are p50 median, with stacked p10 / p90 values after, showing typical ranges.}\par}
\vspace{-2mm}
\end{minipage}
\end{table*}

\section{Evaluation}
\label{evaluation}

We evaluate our system's performance, focusing on the initial construction of the peripheral model and the final rendering quality. All experiments were conducted on a workstation with an Intel i9-13900K CPU and an NVIDIA GeForce RTX 4090 GPU running Ubuntu 22.04. This includes the viewer, which we ran on the same machine for control and comparability with baselines. Foveated renders were generated at a resolution of 512 by 512 pixels, covering a 20° field of view, consistent with prior work~\cite{franke_vr-splatting_2024}. We use the perceptual metrics LPIPS~\cite{zhang_unreasonable_2018}, alongside masked peak signal-to-noise ratio (PSNR) and structural similarity index measure (SSIM) to evaluate image quality. The mask restricts the metric calculation to areas with non-zero alpha in either the ground-truth or rendered image, avoiding distortion due to large empty regions. To account for the inherent variability in rendering complex medical scans, we report median (p50) values accompanied by p10 and p90 percentiles.

\subsection{Datasets}

We focus specifically on medical volumes, which are ideal for our hybrid approach. They contain fine, clinically relevant anatomical structures that benefit from high-fidelity path tracing, yet are naturally bounded. This allows a peripheral model to be generated rapidly, with the maximum size limited to the human body. The training image sets were specifically created for this evaluation from these volumes. They use randomly sampled camera positions without requiring or using a view selection method as presented in \autoref{ss:peripheralmodeltraining}.

We use two volumes with three transfer functions each, creating a diverse dataset of six scenes previously used~\cite{kleinbeck_multi-layer_2025} to aid comparability.
The "Fullbody" volume, derived from the Total Segmentator dataset~\cite{wasserthal_totalsegmentator_2023}\footnote{TotalSegmentator dataset, id \texttt{s0287}, \url{https://zenodo.org/records/10047292}}, is a body scan containing varied structures. It is cropped to reveal lung and organs and has a resolution of $317\times 163 \times 835$ voxels at $1.5 \times 1.5 \times 1.5$ mm each.
The "Leg" volume is adapted from the TCGA-HNSC dataset~\cite{zuley_cancer_2016}\footnote{TCGA-HNSC dataset, id \texttt{TCGA-CV-A6JU}, \url{https://www.cancerimagingarchive.net/collection/tcga-hnsc/}}. It is a high-resolution scan of a lower body, cropped to a single leg without occlusions but features fine details. The volume has a resolution of $195\times 257 \times 3000$ voxels at $0.7 \times 0.7 \times 0.3$ mm each.

\begin{figure*}[t!]
 \centering %
 \includegraphics[trim={3.5cm 0cm 1.8cm 0cm},clip,width=\linewidth]{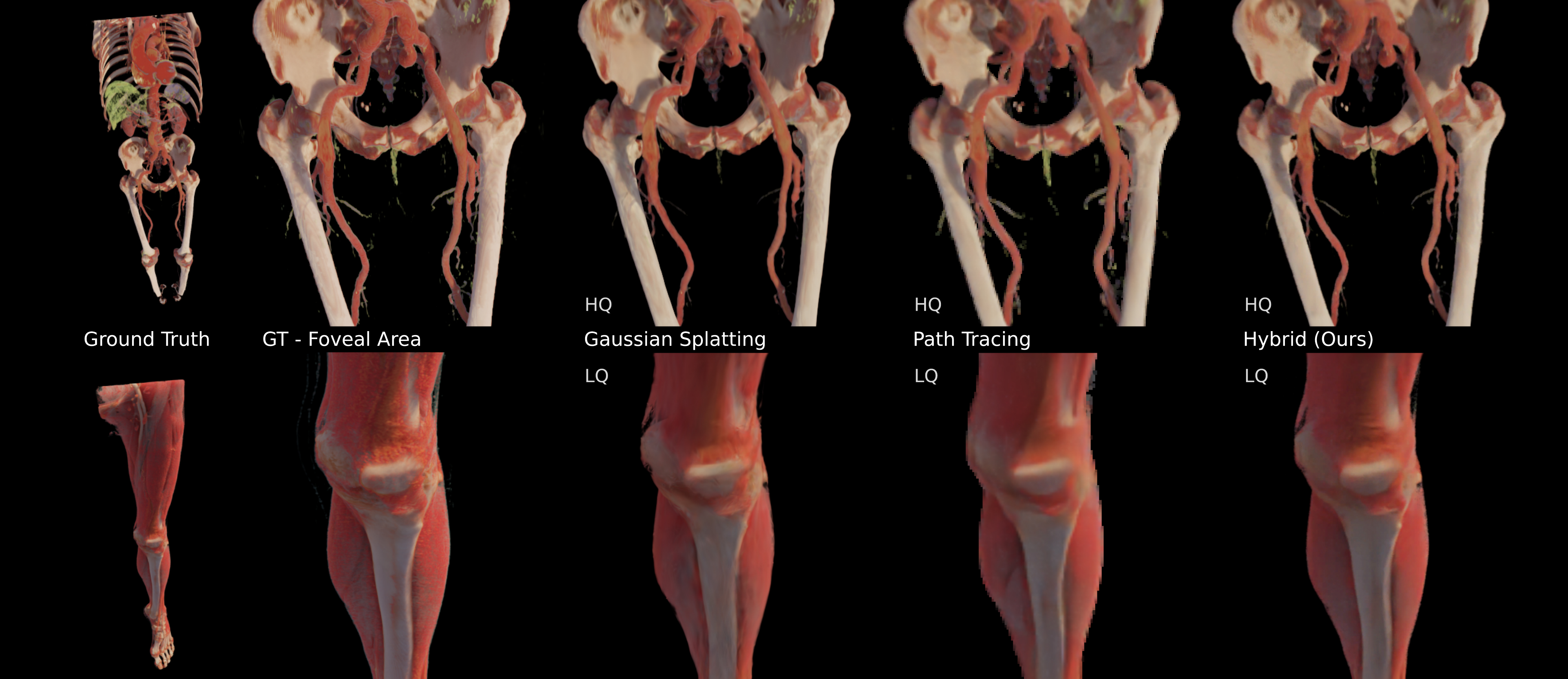}
 \caption{Visual comparison of the resulting quality and baselines (Ground Truth - GT). The upper row shows high-quality settings, as in \autoref{tab:quality-comparison}, the lower row shows lower quality. Our approach retains more detail compared to  \ac{gs} and path tracing, and suffers fewer artifacts.}
 \label{fig:final_compare}
\end{figure*}

\subsection{Peripheral Model}
\subsubsection{Construction}

For peripheral model construction, we create a set of initial images from various views around the target volume. We analyze the trade-off between reconstruction time and quality by varying the number of initial views (8, 12, 16, 32) and samples per pixel (4, 8, 16). For each configuration, we train a peripheral model for 700 iterations using our MiniSplatting2-based~\cite{fang_mini-splatting2_2024} approach (see \autoref{ss:peripheralmodeltraining}).

As shown in \autoref{fig:init_compare}, an initial model can be generated in under 300 ms, though with high error margins. \autoref{fig:init_compare} includes the time taken to render the views, causing the first sample point at 100 iterations to be available at different times. The results indicate that at low render budgets, using more views is more beneficial than increasing the sample count. Just 8 images are often not enough to capture all structures, and 4 \ac{spp} often misses important details. Based on this analysis, we define two configurations for subsequent experiments: a \textit{normal quality} preset (12 views at 8 \ac{spp}) and a \textit{high quality} preset (16 views at 16 \ac{spp}). This allows an initial model to be ready in 300 to 400 ms, with full reconstruction completing in about a second. Using more than 16 views significantly delays the availability of the initial model with diminishing returns. Reported times do not include loading overheads, as we assume these to be implementation-dependent and amortized over multiple images in our system. Increasing the resolution of the training views to 1024 x 1024 increases the time to initial model by about 150 ms for the low quality preset of 12 views at 8 \ac{spp}, and by 280 ms for the high quality preset (16 views x 16 \ac{spp}). Optimization time is increased by approximately 50\%, from ~0.85\,ms to 1.3\,ms for 700 iterations.

Compared to alternatives, our initialization strategy offers superior speed and quality (\autoref{tab:init-compare}). Our method produces a higher-quality model faster than standard \ac{gs} with half the number of Gaussians, as opacity reset struggles to lower Gaussian counts. Other rapid initialization methods that rely on deep learning~\cite{jiang_anysplat_2025, yan_instant_2025} or depth estimation~\cite{chen_mvsplat_2025} are not well-suited for our synthetic data, as they struggle with the transparent or empty regions common in volume rendering.

\begin{table}[t!]
\centering
\caption{Initialization quality on 16 views with 8 \ac{spp} each. Time is reported without loading overhead. Note that for \ac{gs} we did skip densification as with Ours; However, it failed to further reduce Gaussians. Transformer-based models (AnySplat) struggle to predict useful data for synthetically generated anatomy.}
\label{tab:init-compare}
\begin{tabularx}{\columnwidth}{@{}l|YYYYY@{}}

\toprule
            & PSNR                       & MPSNR                      & LPIPS                   & Gaussians            &Time \\ \midrule
3DGS        & \pval{26.24}{28.30}{30.38} & \pval{19.98}{21.95}{25.07} & \pval{.081}{.119}{.155} & \pval{20}{20}{20} & \pval{1.02}{1.05}{1.09} \\[2pt] % Increased vertical space between rows
AnySplat    & \pval{13.82}{14.90}{15.33} & \pval{12.45}{13.73}{14.08} & \pval{.379}{.431}{.461} & \pval{2990}{3006}{3027} & \pval{1.07}{1.08}{1.08} \\[2pt]
Ours                                & \pval{26.10}{28.85}{31.06} & \pval{18.75}{22.60}{25.51} & \pval{.079}{.108}{.144}      & \pval{9}{10}{11} & \pval{0.89}{0.89}{0.93} \\ \bottomrule
\end{tabularx}
\begin{minipage}{\columnwidth}
\vspace{2mm}
{\raggedright \scriptsize{\textit{Note:} MPSNR denotes masked PSNR values. Values are p50 median, with stacked p10 / p90 values after, showing typical ranges. Gaussians in are reported in thousands.}\par}
\vspace{-2mm}
\end{minipage}
\end{table}

\subsubsection{Continual Training}

We further evaluate the model's improvement over time by refining it with additional views, simulating prolonged user interaction. We augment the initial models with larger sets of 64 and 512 closer-up images, representing prolonged use and maximum quality. The model optimizes for up to 4000 iterations, including 400 iterations of densification to capture finer details. After these iterations, we empirically found only very little further model improvement, with improvement after 2000 iterations already being limited.

The results in \autoref{tab:ext-compare} show a substantial increase in visual quality as the model is refined, with the Gaussian count growing to between 20k and 50k. The training proceeds at over 500 iterations per second, allowing a model with relevant improvements to be sent to the viewer every few seconds. We note that improvements become marginal with very large image sets. 

When using foveated and subsequently undistorted views for training, as would be generated during use with higher quality targets, the quality metrics decrease by 5-10\%, depending on the anatomical scene. A similar decrease happens for both images with foveation fixed at the image center, as well as images with variable foveation points, with less than 1\% quality difference between them.

\subsection{Rendering Analysis}

\begin{table*}[]
\centering
\caption{Comparison of foveated path tracing with the peripheral \ac{gs} model with either system alone. Our approach outperforms both real-time path tracing and precomputed \ac{gs} in the peripheral region, while offering competitive visuals in the full image, focusing quality where it matters. Note that LPIPS worsens in foveated views, as it takes the ample blank space in full views into account, unlike masked metrics.}
\label{tab:quality-comparison}
\begin{tabularx}{\linewidth}{@{}l|YYYYYYYY@{}}
\toprule
&  \multicolumn{4}{c}{Baseline} & \multicolumn{4}{c}{Ours} \\ 
\cmidrule(l){2-5} \cmidrule(l){6-9} &  PT &  PT\,HQ &  3DGS &  3DGS\,HQ & Ours & Ours\,+v & Ours\,HQ & Ours\,+v\,HQ \\ \midrule
Preparation (s)   & \pval{}{-}{}         & \pval{}{-}{}            & \pval{52.6}{60.1}{71.7} & \pval{107.2}{124.7}{144.6} & \pval{1.0}{1.0}{1.0} & \pval{1.0}{1.0}{1.0} & \pval{1.1}{1.1}{1.2} & \pval{1.1}{1.1}{1.2} \\
Render (ms)       & \pval{5.5}{5.6}{6.0} & \pval{16.2}{19.1}{19.5} & \pval{}{<0.5}{} & \pval{}{<0.5}{} & \pval{5.6}{5.9}{6.1} & \pval{5.6}{5.9}{6.1} & \pval{15.6}{17.2}{18.8} & \pval{15.6}{17.2}{18.8} \\ \midrule
\multicolumn{9}{l}{\kern-2mm\textbf{Full image}} \\  
~~MPSNR ↑         & \pval{18.62}{22.38}{25.22} & \pval{20.60}{24.66}{27.59} & \pval{18.47}{19.58}{22.09} & \pval{21.67}{21.88}{24.60} & \pval{17.18}{19.46}{22.50} & \pval{21.92}{22.31}{25.35} & \pval{17.29}{19.61}{22.65} & \pval{21.93}{22.30}{25.38} \\
~~MSSIM ↑         & \pval{.649}{.739}{.786}   & \pval{.737}{.797}{.828}   & \pval{.650}{.723}{.758}   & \pval{.792}{.828}{.863}   & \pval{.625}{.728}{.759}   & \pval{.763}{.801}{.839}   & \pval{.644}{.739}{.769}   & \pval{.755}{.815}{.850} \\
~~LPIPS ↓         & \pval{.071}{.089}{.109}   & \pval{.058}{.078}{.092}   & \pval{.056}{.080}{.094}   & \pval{.036}{.061}{.074}   & \pval{.068}{.084}{.106}   & \pval{.051}{.075}{.091}   & \pval{.066}{.082}{.103}   & \pval{.048}{.073}{.087} \\  \midrule
\multicolumn{9}{l}{\kern-2mm\textbf{Foveated region (20°)}} \\  
~~MPSNR ↑         & \pval{18.60}{22.48}{26.64} & \pval{20.54}{24.79}{28.93} & \pval{19.24}{20.13}{23.92} & \pval{21.66}{22.72}{26.24} & \pval{22.36}{23.31}{27.80} & \pval{23.17}{24.06}{28.14} & \pval{23.16}{23.87}{28.55} & \pval{23.87}{24.91}{28.90} \\
~~MSSIM ↑         & \pval{.668}{.716}{.783} & \pval{.766}{.788}{.830} & \pval{.639}{.705}{.780} & \pval{.750}{.815}{.885} & \pval{.726}{.780}{.839} & \pval{.740}{.803}{.854} & \pval{.764}{.819}{.866} & \pval{.775}{.833}{.880}   \\
~~LPIPS ↓         & \pval{.158}{.195}{.230} & \pval{.122}{.159}{.182} & \pval{.118}{.169}{.212} & \pval{.072}{.120}{.154} & \pval{.114}{.147}{.176} & \pval{.110}{.147}{.176} & \pval{.094}{.123}{.146} & \pval{.090}{.121}{.142}    \\ \bottomrule
\end{tabularx}
\begin{minipage}{\linewidth}
\vspace{2mm}
{\raggedright \scriptsize{\textit{Note:} MPSNR and MSSIM denotes masked PSNR / SSIM values. Values are p50 median, with stacked p10 / p90 values after, showing typical ranges. Rendering a 2000 x 2000 px texture. Path Tracing (PT) with 2 \ac{spp} or 6 \ac{spp} (HQ) upscaled, both with foveated rendering. \ac{gs} trained with same 512px 4spp images as Ours, or 1024px 128spp images. Ours with 4 \ac{spp} or 16 \ac{spp} (HQ). Peripheral training with additional views indicated with +v if applicable.}\par}
\vspace{-2mm}
\end{minipage}
\end{table*}

We compare our hybrid pipeline with standalone \ac{gs} and foveated path tracing. We compare to normal and higher quality path tracing (both with foveated rendering), targeting similar frame times as our approach, to show how the addition of the peripheral model influences visual results. The comparison to just \ac{gs} serves as a baseline of what would be possible with more rendering and training time spent, without requiring heavy real-time path tracing. It does incur 1 to 2 minutes of setup time with our settings, compared to under a second for the initial model with our approach. Our results indicate improved performance compared to its components. It delivers superior foveal quality compared to real-time path tracing. At the same time, it is competitive with \ac{gs} at a fraction of the setup time required. \autoref{tab:quality-comparison} presents a quantitative comparison for both the full view and the foveated region, while \autoref{fig:final_compare} offers a visual comparison. We compare both our normal quality preset, suitable for real-time rendering, and our high quality preset, utilizing reprojection to enable improved quality. Note that our rendering time is dominated by rendering the path traced image; reprojection and peripheral model rendering in the viewer happen in under a millisecond on the desktop Linux machine, similar to the 3DGS baseline. Similarly, our method usually renders at above 72 FPS standalone on a Meta Quest 3 \ac{hmd}, building on previous work with heavier models\cite{kleinbeck_multi-layer_2025}.

% Quality
In the foveal region, our method outperforms the baseline path tracer when benchmarked at similar frame times, especially on the perceptual metric. Note that while the masked metrics slightly increase in the foveated region, LPIPS decreases, due to them taking the ample empty space of full views into account. Comparing our two quality levels shows that the quality improvement from more samples is limited in peripheral regions and more pronounced in the foveated view. Continual training primarily improves the peripheral model, with only a minor impact on the foveated area, which is dominated by the path-traced image. An empirical adjustment in the viewer to increase splat scale and opacity by 10\% yielded a 2-3\% improvement in overall metrics.

% Speed
Our rendering settings are tuned to achieve frame times suitable for \ac{vr}, and further adjustments are straightforward by adjusting render parameters like resolution and \ac{spp}. Since our method and the path tracing baseline share the same experimental volume renderer, their performance is directly comparable. Denoising (2\,ms) and albedo calculation (1\,ms) contribute significantly to the frame times. These times are incurred on the machine doing the volume rendering. In the viewer, rendering the peripheral \ac{gs} model was CPU-bound on our desktop machine, taking less than 0.5 ms. Consistent with previous work~\cite{kleinbeck_multi-layer_2025}, these small models can be rendered well in under 10 ms on mobile \ac{vr} headsets like the Meta Quest. Mobile performance is very variable and viewpoint-dependent due to fill-rate limitations, making it hard to give exact numbers.

%Notes
The file size of the peripheral model is typically a few megabytes after simple quantization, making it suitable for real-time transmission. For this evaluation, all path-traced views, for both our method and the baseline, use fixed foveated rendering. This was done for simplicity in testing and evaluation, the approach itself supports variable foveation points. Our path tracer is a research codebase based on recent work~\cite{hofmann_efficient_2021}, tuned for this application.

\section{Discussion}

Our hybrid approach combines streamed foveated path tracing with rapidly generated Gaussians for the periphery. It yields superior results in both speed and visual quality compared to standalone path tracing or \ac{gs} in the foveal region. Peripheral Gaussian regeneration is fast enough for interactive use at a presentable visual quality, as highlighted in \autoref{fig:init_compare}. The approach benefits from advances in path tracing, denoising, and peripheral reconstruction: as any of these components improves, so does our approach. Relative to streaming the entire anatomy, our approach improves foveal quality, reduces the risk of disocclusion by continuously rendering the periphery, improves latency masking via depth-guided reprojection, and lets users trade rendering speed for quality by minimizing the reprojected area. Compared to rendering only Gaussians, our method removes lengthy precomputation, improves visual fidelity, and supports arbitrary zoom and interaction by regenerating the periphery on demand. That said, the method requires more computing than standalone \ac{gs}; this extra work is not done in the viewer, preserving the ability to use mobile \ac{vr} for high-quality rendering. Generally, GPU workloads are not evenly balanced. Peripheral-model creation and refinement are computationally expensive, but not ongoing: in our experiments, they required relatively few iterations and images to improve quality. The system can run on one to three GPUs with trade-offs: colocating peripheral generation and path tracing on the same GPU simplifies the setup but delays model updates and slightly degrades path-tracing throughput.

We strived for comparable quality in the peripheral region compared to just path tracing, while improving visuals in the foveated region. It should be kept in mind that with path tracing, the quality differential between these two areas can be shifted arbitrarily by adjusting the foveation strength. On the other hand, we argue that our approach is expected to improve upon only path tracing, by virtue of having to shade fewer pixels. Looking at the metrics of only \ac{gs} indicates that further quality improvements to our peripheral model are possible with less aggressive simplification, at the expense of some render speed reduction in the viewer. The metrics, however, do not capture all perceptual effects. Color and luminance discrepancies at the boundary between foveal textures and peripheral Gaussians are more noticeable to human observers. These boundary artifacts are most apparent during initialization with few views and low \ac{spp}. To mitigate this, we recommend using at least 12 initial images and boosting the splat scale and opacity on the viewer side. This also improves visibility of small, semi-transparent structures (e.g., vessels) that the reconstruction tends to underrepresent. These issues could be mitigated during model training by adding luminosity-aware loss terms or dilation. Further perceptual enhancements like increased vibrancy/contrast or extending the noise texture of foveal renderings into the peripheral Gaussians could improve transitions and peripheral quality~\cite{patney_towards_2016}.

Several simplifying design choices must be considered when generalizing our results. We used a foveated resolution of $512\times512$, matching previous works on mobile \ac{vr}~\cite{franke_vr-splatting_2024}. Higher resolutions are possible at the cost of render time and may be more valuable than additional samples per pixel. Textures can also be upsampled as necessary. We used a simple TCP messaging system and restart the continual training process when a sufficient set of new images is available; to simulate continued use, we also evaluated with a pre-created image set. Although pose distributions for refinement images in practice will be different and often more concentrated, our experiments show the potential achievable quality. With highly focused gaze behavior, the periphery may improve slowly in off-focus regions, but typical human scanning and saccades should still drive broad reconstruction in regions of interest. As we specifically created test image sets, the view selection and its hyperparameters were not extensively tested. Better strategies can likely be found with focused tests, however we want to stress the general need for some kind of view selection to manage resources.

There are additional opportunities to increase system utility. Lightweight view-selection algorithms could raise initial quality, but runtime cost is critical. Our pipeline uses only 10 to 20\,ms per view, so simple strategies that avoid expensive computation may be preferable to a complex selection that competes with rendering time. View selection could also be used during continual refinement to prioritize underexplored regions~\cite{niedermayr_application_2024}. The \ac{gs} optimizer could pause when gains are negligible and instead request new views. Persisting trained peripheral models across sessions could accelerate initial quality for repeat sessions. Likewise, sharing optimization state between viewers would improve multi-user scenarios. Finally, the rendering adjustments from StopThePop~\cite{radl_stopthepop_2024} and Optimal Projection~\cite{huang_error_2025} were not included for speed and simplicity. We found that these issues were not prominent with our smoothly rendered anatomical data, which is mostly drawn in the scene center, but they could be valuable for wider user bases. 

\subsection{Limitations}

Our work is not without limitations. We do not model the full system interconnections, concurrent works cover streaming of foveated rendering and \ac{gs} clouds. Our focus is on evaluating the individual components that make up the system. The system as implemented is functional but not in a state to run end-to-end tests with users. This means it is not yet possible to make statements on how the system would feel for users during use in realistic gaze scenarios. The view selection algorithm is a simple way to handle resource constraints during continued use, but not thoroughly evaluated. Real user gaze behavior on medical volumes would be needed for comprehensive analysis. Compared to pure path tracing or static \ac{gs}, the hybrid method can introduce visible artifacts at the boundaries of depth-projected textures despite blending. Fast head movements can expose reprojection failures, particularly during close-up views. In our testing, these artifacts were not common during typical use, but they remain an issue. Peripheral model construction can undersample details, making them appear smaller or more transparent and producing perceptually darker peripheral models in some cases. Finally, the approach scales best for scenes of limited spatial extent, like anatomical objects. We expect that in large virtual environments, the time required to generate a peripheral model of sufficient quality and render it at interactive rates grows quickly, limiting applicability.

\subsection{Future Work}

The components of our approach offer various options for further work to refine:  Initial point clouds could be created by directly estimating \ac{gs} parameters like scale or rotation from rendered views, extending existing works with alpha channel awareness, speeding up time to initial peripheral model. Continual training can benefit from localized refinement strategies and stronger heuristics~\cite{ackermann_cl-splats_2025}. Reprojection and disocclusion handling should be made more robust for close-up viewing, potentially by strategically filling holes with information from the peripheral cloud. Future work could also focus on improving the perceptual experience of the \ac{gs} model by tuning it for the human peripheral vision. Training and viewing time adjustments to contrast or luminosity, as well as other parameters, could increase perceptual quality and help users orient themselves. Improvements to the system interconnects would enable user studies: These could investigate technical improvements such as improved view selection and model training in response to real gaze and free foveation as well as research into the perceptual benefits of the hybrid approach and quantify parameters for trade-offs like foveal image quality vs. update frequency.

\section{Conclusion}
We present a hybrid rendering system that combines foveated path tracing with a peripheral Gaussian Splatting model for interactive medical volume visualization. This approach successfully bridges the gap between the high fidelity of path tracing and the real-time performance of pre-computed representations. Key contributions are a method for rapidly regenerating the peripheral model in seconds, as well as a reprojection strategy for inserting the path-traced views into the peripheral cloud. This enables interactive exploration and adjustments to the anatomy and effectively masks latency. Compared to established approaches, we demonstrate improved perceptual quality in the foveal region at similar rendering speeds. Thus, our work lays the groundwork for interactive, high-fidelity immersive visualization for medical data exploration on mobile \ac{vr} devices.

\section*{Supplemental Materials}
\label{sec:supplemental_materials}

Supplementary material is publicly available at \url{https://osf.io/6thdu/}. Source code of the path tracer, training, and Unity renderer is available at \url{https://github.com/roth-hex-lab/Hybrid-Foveated-Path-Tracing}. The original CT volumes used in this work are available from their respective authors.

\acknowledgments{
\label{sec:figure_credits}

The text of this work was proofread and spell-checked, and code used in this publication has been partially written or adjusted by large language models (OpenAI ChatGPT, Google Gemini). All proposed changes have been checked for correctness by the authors, and all text bodies have been written by the authors. All figures and other media were created solely by the authors.
}

\bibliographystyle{abbrv-doi-hyperref-narrow}

\bibliography{literature}
\end{document}